\documentclass[twocolumn,prl,showpacs,preprintnumbers]{revtex4}
\usepackage{epsfig}
\usepackage{amssymb}

\newcommand{\Z}{{Z \!\!\! Z}}
\newcommand{\beqn}{\begin{eqnarray}}
\newcommand{\eeqn}{\end{eqnarray}}
\newcommand{\eq}[1]{(\ref{#1})}

\newcommand{\dd}{\mbox{d}}
\newcommand{\dD}{{\cal D}}
\newcommand{\dual}{\mbox{}^{\ast}}

\newcommand{\cint}[1]{\!\int\nolimits^\pi_{-\pi} \!\!\!\dD #1}
\newcommand{\iint}[1]{\!\int\nolimits^\infty_{-\infty} \!\!\!\dD #1}
\newcommand{\nsum}[2]{\sum_{#1 \in \Z(c_#2)}}
\newcommand{\ClosedSum}[2]{\sum_{\stackrel{#1 \in \Z(c_#2)}{\delta #1 = 0}}}
\newcommand{\DualClosedSum}[2]{\sum_{\stackrel{\dual #1 \in \Z(\dual c_#2)}{
                               \delta \dual #1=0}}}

\newcommand{\Npoly}{N_{\mathit{poly}}}

\begin{document}

\title{Screening and confinement in $U(1)^{N-1}$  Abelian
effective theories}

\author{Tsuneo~Suzuki$^1$ and M.~N.~Chernodub$^{1,2}$}
\affiliation{
$^{1}$Institute for Theoretical Physics, Kanazawa
University, Kanazawa 920-1192, Japan\\
$^{2}$Institute of Theoretical and  Experimental Physics,
ITEP,
B. Cheremushkinskaja 25, Moscow, 117259, Russia}

\preprint{KANAZAWA-02-15}
\preprint{ITEP-LAT/2002-06}

\begin{abstract}
We discuss the effect of dynamical charged particles in
$U(1)^{N-1}$ Abelian effective theories of QCD.  Screening and
confinement at zero temperature are explained qualitatively. The
effect of dynamically charged particle is expressed in terms of an
effective action of integer electric currents. When a screening is
expected, the static potential shows a flattening in the
long-range region and a linear behavior in the intermediate
region. We show why the screening is better observed in the
Polyakov-loop correlators than in the Wilson loops. The breaking
of the adjoint string is explained without the Z(N) picture.
\end{abstract}

\pacs{12.38.Aw,11.15.Ha,14.80.Hv}

\date{\today}

\maketitle

Understanding of non-perturbative aspects of strongly coupled
field theories -- such as QCD -- is always a challenging task.
From the study of confinement in QCD, it has been known that
duality and topology are two important keywords. Namely 't~Hooft
idea~\cite{'tHooft:1981ht} of monopole condensation after Abelian
projection to $U(1)^{N-1}$ of $SU(N)$ QCD has been shown to be
successful in the infrared region. Abelian and monopole components
seem to be dominant in the long-range region of quenched
QCD~\cite{suzuki90}. An infrared effective monopole action has
been derived in the continuum limit after a block-spin
transformation of monopole currents~\cite{shiba95,nakamura}. It is
a quantum perfect action described by monopole currents. The
monopole action shows that monopole entropy dominates over the
energy. That is, monopole condensation occurs in the confinement
phase. Using a quantum perfect operator~\cite{fujimoto}, we can
evaluate the string tension analytically and prove the continuum
rotational invariance of the static potential~\cite{chernodub00}.

Although the 't~Hooft scenario seems correct in many respects,
there remains an unsolved serious problem even in quenched QCD and
also in full QCD. It is the screening-confinement problem
extensively discussed in recent
publications~\cite{StringBreaking}. How can screening be explained
in this scheme? Consider a static pair of charge 2 particles in
quenched QCD. Since off-diagonal gluons can screen the charge 2
particle, the static potential is expected to show flattening in
the long-range region, i.e., screening. Actually, an adjoint
Wilson loop and a charge 2 Wilson loop show such behaviors in the
strong coupling region. This problem was discussed extensively in
Ref.\cite{greensite}. The theory in terms of Abelian link fields
or Abelian monopole currents alone becomes highly non-local if we
integrate out off-diagonal gluon fields after an Abelian
projection. Namely  we have to keep all charge 2 Abelian Wilson
loops in the effective action written by Abelian link fields to
reproduce the screening of charge 2. Needless to say, such an
Abelian effective action is useless. The same problem is more
serious in the real full QCD, since a fundamental charge also is
screened. The authors in Ref.\cite{greensite} suggest that the
relevant quantity in the confinement mechanism is not the Abelian
monopoles but the $Z(N)$ center-vortices which can explain the
screening problem~\cite{ZNpicture}.

The aim of this note is to show how the screening and confinement
problem is solved qualitatively
in the framework of $U(1)^{N-1}$  Abelian dynamics. We find
the effect of dynamical charged particles can be described
in terms of integer  electric currents on the lattice.

Let us first review the case without a dynamical char\-ged particle.
For simplicity, we consider only $SU(2)$ QCD  $\to U(1)$ and use the
differential forms on the lattice~\cite{DiffLattice}.

Consider a modified compact QED action of the form
\beqn Z_1 =
\cint{\theta} \!\!\!\nsum{n}{2} \!\!\!
e^{ - \frac{1}{4\pi^2}({\mathrm d} \theta + 2 \pi n, \Delta D
({\mathrm d} \theta + 2 \pi n)) + i (Q J,\theta)}\,,
\label{Z1}
\eeqn
where the operator $D$ is a general differential operator,
$Q$ is a charge of an external source and
the current $J$ takes $\pm 1$ along the Wilson loop.
$\Delta$  is the  Laplacian on the lattice.
Following Ref.~\cite{chernodub00}
one can reduce the partition function~\eq{Z1}
to that of the following monopole action
\beqn
Z_2 = \!\!\! \DualClosedSum{k}{1}e^{- (\dual k, D\dual k)+2\pi i(\dual k,
\dual N) -\pi^2Q^2(J,(\Delta^2D)^{-1}J)}\,,
\label{Z2}
\eeqn
where $\dual N\equiv\Delta^{-1}\delta\dual S$ and $S$ is an open
surface spanned on the  Wilson loop, i.e., $\delta S\equiv QJ$.
Here and below the constant pre-factors are ignored.
Numerically it is  found~\cite{nakamura}  that the  Abelian effective monopole
action derived from $SU(2)$ QCD can be described well in
terms of  two-point interactions alone in the long-range
region as in (\ref{Z2}).

It is also possible  to transform the monopole partition
function (\ref{Z2}) into the string representation as follows~\cite{chernodub00}:
\beqn
Z_3
= \sum_{\stackrel{\sigma \in \Z(c_2)}{\delta\sigma=QJ}}e^{-\pi^2(\sigma,
(\Delta D)^{-1}\sigma)}\,,
\label{Z3}
\eeqn
where the summation goes over the surfaces spanned ($|Q|$ times)
on the external current $J$.

In the infrared limit the operator $D$ is approximated well by
Coulomb+self+nearest neighbor terms~\cite{nakamura}, that is,
$D=\beta\Delta^{-1}+\alpha+\gamma\Delta$, where $\beta$, $\alpha$
and $\gamma$ are renormalized coupling constants of the monopole
action.  They satisfy the
relation $\beta\gg\alpha, \gamma$. Then
the Wilson loop can be estimated from Eq.\eq{Z3}:
\beqn
{\langle W_Q(R,T) \rangle}= {\mathrm{const.}}\,
e^{ - \kappa\, |Q| \, RT + \cdots}\,,
\label{string:mon}
\eeqn
where $RT$ is the area of the minimal surface spanned on the
contour $J$. There are $|Q|$ such surfaces which must be parallel
to each other~\cite{Faber} to maximize the contribution for $|Q|
\geqslant 2$. This explains linearity in $|Q|$ in
Eq.~\eq{string:mon}.
When $\beta\gg\alpha, \gamma$,  the string
tension is evaluated approximately as $\kappa = \pi^2 \slash \beta$.

Now let us introduce a dynamical charged particle. A charged
vector field corresponding to gluon fields and a fermion field
corresponding to quark fields are rather complicated to deal with.
Hence we consider first a charged scalar Higgs field in the London
limit as a simple example. The radial part $\rho_x$ of the scalar
Higgs field $\Phi_x = \rho_x \, e^{i \vartheta_x}$ is frozen and the
dynamical variable is the compact Higgs phase $\vartheta \in
[-\pi,\pi)$ which carries the electric charge $q$. The Higgs field
action is written in the Villain representation:
\beqn
Z_4[\theta] = \cint{\vartheta}  \nsum{l}{1}
e^{ -({\mathrm d} \vartheta + q \theta + 2 \pi l, G
({\mathrm d} \vartheta + q \theta + 2 \pi l))} \,,
\label{Z4}
\eeqn
where $G$ is a local operator.

The integration of the phase of the Higgs field $\vartheta$ can be
represented as a weighted sum of the Wilson loop over the closed
charged currents following Ref.~\cite{EihornSavit}:
\beqn
Z_4 & = &
\cint{\vartheta}  \iint{F} \!\! \nsum{l}{1} \!\! e^{-
(F, (4G)^{-1}F)+ i
(F, {\mathrm d} \vartheta + q \theta + 2 \pi l)} \nonumber\\
& = & \cint{\vartheta} \nsum{j}{1} e^{-(j, (4G)^{-1} j) + i
(j, {\mathrm d} \vartheta + q \theta)}\\
&= &\ClosedSum{j}{1} e^{-(j, (4G)^{-1} j)+i(qj,\theta)} \,,
\label{Higgs:trajectories}
\eeqn
where we consecutively introduce the Gaussian integration over the
non-compact link variable $F$, apply the Poisson summation
formula, $\sum_l e^{2 \pi i (l,F)} = \sum_j \delta(j-F)$,
integrate over the fields $F$ and $\vartheta$. The last integration
gives the closeness constraint $\delta j = 0$.

Hence the Villain type compact QED
with the charged scalar field is written as
\beqn
Z_5 &= &\ClosedSum{j}{1}e^{-(j, (4G)^{-1} j)}\cint{\theta}
\!\!\nsum{n}{2} \label{Z5} \\
& & e^{ - \frac{1}{4\pi^2}({\mathrm d} \theta + 2 \pi n, \Delta D
({\mathrm d} \theta + 2 \pi n)) +i(QJ+qj,\theta)}\,. \nonumber
\eeqn
This expression can be reduced further to the monopole-electric
and the string-electric current models:
\beqn
Z_6&=&\hspace{-.4cm}\ClosedSum{j}{1}\DualClosedSum{k}{1}
e^{-(\dual k, D\dual k)+2\pi i(\dual k,
\dual N+\dual n)}\nonumber \\
& & \times e^{-(j,(4G)^{-1}j)-\pi^2(QJ+qj,(\Delta^2D)^{-1}(QJ+qj))}
\label{Z61}\nonumber\\
&=&\ClosedSum{j}{1}\sum_{\stackrel{\sigma \in \Z(c_2)}{\delta\sigma=QJ+qj}}
\hspace{-.3cm}e^{-(j,(4G)^{-1}j)
-\pi^2(\sigma, (\Delta D)^{-1}(\sigma))}\,,
\label{Z6}
\eeqn
where $\dual n\equiv\Delta^{-1}\delta\dual s$ and
$s$ is a surface spanned $q$ times on the dynamical current $j$,
{\it i.e.}, $\delta s\equiv qj$.

{}From Eq.(\ref{Z6}), we can see how the
screening-con\-fi\-ne\-ment problem is solved. If $Q/q \notin \Z$,
 the static potential can not be screened completely. For
example, in the case of $Q=1$ and $q=2$, the leading string
tension is equivalent to the string tension without presence of
the dynamical charges as in (\ref{string:mon}). In this case,
$j=0$ in the sum over $j$ gives the leading term while other terms
coming from non-zero $j$ show stronger damping.

On the other hand, if $Q/q \equiv N \in \Z$ (as in the
case of $Q=1$ and $q=1$) then the expectation value of the
$R\times T$ Wilson loop is expanded as a perimeter term ($\propto
(R+T)$) given by the action of the dynamical screening
current with $QJ+qj=0$ and the area-law term:
\beqn
{\langle W_Q(R, T) \rangle} & = & c_0 \, e^{-m N^2 (R+T)}
\label{string:all} \\
& & + c_1 \, e^{- \kappa q R T - m (N-1)^2 (R+T)}  + \cdots\,.
\nonumber
\eeqn
These terms correspond to the screening currents, respectively,
$j = - N J$ and $j = (- N + 1) J$. The string breaking is seen at
large distances. It is important to note that the area-law behavior is
observed in the intermediate region even when the screening occurs.

One can easily show why the screening is better observed in the
case of the Polyakov loop correlator rather than in the case of
the Wilson loop quantum average
as found recently~\cite{StringBreaking}. Consider two Polyakov loops
separated by a distance $R$ corresponding to a pair of static
quark and anti-quark in the case of periodic boundary condition in
the time direction. The two leading terms in the average are:
$$
{\langle P_Q(0)P_{\bar Q}(R) \rangle} = d_0 \, e^{-m N^2 T}
+ d_1 \, e^{- \kappa q R T - m (N-1)^2 T}+ \cdots\,.
$$
A comparison of this expression with Eq.\eq{string:all} gives that
the area-law terms are the same for both averages.
However, the perimeter terms are different: for the
Wilson loop the perimeter term contains additional suppression
factor $\sim e^{- m (2N - 1) R}$ with respect to the area term.
Thus at sufficiently large separations between the sources the
perimeter term in the Wilson loop average may not be found
numerically. As a result, the string breaking may not be
observed in the Wilson loop even if the breaking can be seen
in the Polyakov loop correlator.

To derive numerically such an  effective action (\ref{Z61}) in
terms of monopole and electric currents is very interesting.
Integer monopoles can be defined  following
DeGrand-Toussaint\cite{degrand} while definition of the integer
electric currents in terms of original Abelian link fields seems
to be impossible. Nevertheless, a given effective matter action as
in (\ref{Z4}), the action in terms of integer electric currents
(\ref{Z61}) can also be derived.

Now we consider the case of dynamical charged link variables which
correspond to off-diagonal gluon fields after the Abelian
projection of QCD. The SU(2) link variable can be parameterized as
$U_{x,\mu} = c_{x,\mu} u_{x,\mu}$ where the matrix $c_{x,\mu}$
corresponds to the off-diagonal gluon while $u_{x,\mu}$ represents
the diagonal photon contribution: $c^{11}_{x,\mu} = c^{22}_{x,\mu}
= \cos \eta_{x,\mu}$, $c^{12}_{x,\mu} = c^{21 *}_{x,\mu} = i \sin
\eta_{x,\mu} e^{i\varphi_{x,\mu}}$, $u^{11}_{x,\mu} = u^{22
*}_{x,\mu} = e^{i\theta_{x,\mu}}$ and $u^{12}_{x,\mu} =
u^{21}_{x,\mu} = 0$. The link fields are restricted, $\eta \in
[0,\pi/2]$, $\theta,\varphi \in [-\pi, \pi)$.

The meaning of these fields can be understood after applying the
$U(1)$ gauge transformations, which are allowed in an Abelian
projection. The field $\theta$ is the Abelian gauge field and
$\varphi$ is the vector field (phase of the off--diagonal gluon
field) with the electric charge two, respectively: $\theta_{x,\mu}
\to \theta_{x,\mu} +\xi_x -\xi_{x+\hat{\mu}}$, $\varphi_{x,\mu}
\to \varphi_{x,\mu} +2 \xi_x$. In the continuum limit these
transformations reduce to $\theta_\mu(x) \to \theta_\mu(x) -
\partial_\mu \xi(x)$ and $\varphi_\mu(x) \to \varphi_\mu(x) + 2
\xi(x)$, respectively. The variable $\eta_{x,\mu}$ is $U(1)$ gauge
invariant.

The SU(2) plaquette action  ${\mathrm{Tr} U_P}$ in terms of the
angles $\eta, \ \theta $ and $\varphi$ can be written as a sum of
three parts~\cite{Chernodub:pw}. The first part is proportional to
the compact QED action, $S^\theta  = - \beta_{SU(2)} \cos \dd
\theta\, \prod^4_{i=1} \cos\eta_i$, where $\beta_{SU(2)}$ is the
SU(2) gauge coupling and the subscripts $1,...,4$ correspond to
the standard notations of the plaquette links:  $1 \rightarrow
\{x,x+\hat{\mu}\},...,4 \rightarrow \{x,x+\hat{\nu}\}$. The second
part, $S^{\theta\varphi}$, contains 6 terms representing the
interaction of the matter field $\varphi$ with the gauge field
$\theta$. All these terms are proportional to the products
$\cos\eta_1\, \cos\eta_2\, \sin\eta_3\, \sin\eta_4$ and the like.
The last part of the action is given by $S^\varphi  = - \beta_{SU(2)} \cos
\tilde{\dd} (\varphi+\theta) \, \prod^4_{i=1} \sin\eta_i$, where
$\tilde{\dd} \varphi = \varphi_1 - \varphi_2 + \varphi_3 -
\varphi_4$.

The simplest way to demonstrate the adjoint string breaking is to
use a mean--field approximation for the field $\eta$ (for our
purposes the fluctuations of this field are not essential). We set
$\cos\eta_{x,\mu} \to \langle
\cos\eta_{x,\mu} \rangle = c$ and $\sin\eta_{x,\mu} \to \langle
\sin\eta_{x,\mu} \rangle = s$, respectively. Then the
self--interactions of the gauge and matter fields can be written
in a short form, $S^\theta = - \beta_{SU(2)} c^4 \cos \dd \theta$
and $S^\varphi = - \beta_{SU(2)} s^4 \cos \tilde{\dd} (\varphi +
\theta)$, respectively.

The role of the actions $S^{\theta\varphi}$ and $S^{\varphi}$ can
be elucidated by noticing that they basically contains the
following combinations of the fields: self--interaction of the
gauge field, $\dd \theta$, the interaction of the gauge field with
the off--diagonal field, $H_{x,\mu\nu} \equiv 2 \theta_{x,\mu} +
\varphi_{x,\nu}- \varphi_{x+\hat\mu,\nu}$ and self--interaction of
the off-diagonal field, $C_{x,\mu\nu} \equiv \varphi_{x,\mu} -
\varphi_{x,\nu}$. Thus from the bare Wilson gauge action we get the
following mean-field types of interactions involving the
off-diagonal fields:
$$
S^{\theta\varphi}_{x,\mu\nu}+S^{\varphi}_{x,\mu\nu}\longrightarrow
\cos H_{x,\mu\nu} + \cos H_{x,\nu\mu} + \cos C_{x,\mu\nu}\,.
$$
The tensor $H_{\mu\nu}$ is not antisymmetric contrary to $C_{\mu\nu}$.
Therefore we need two terms with $H$ at each plaquette.

A general effective action of charged link variables can be written in
terms of two periodic actions:
\beqn
{\tilde S}^{\theta\varphi}_{x,\mu\nu} = F_H(H)+F_H(H')+F_C(C)\,,
\label{S:th:phi}
\eeqn
where the form $H \equiv H_{x,\mu\nu}$ ($H' \equiv H_{x,\nu\mu}$)
corresponds to the positively (negatively) oriented plaquette
interactions of the matter and gauge fields.
Now let us  consider the adjoint
string breaking in the $SU(2)$ gluodynamics using
the following partition function:
\beqn
Z_{7} = \cint{\theta} \cint{\varphi} e^{
- \sum_P \bigl(S^{\theta}_P(\theta) + {\tilde
S}^{\theta\varphi}_P(\theta,\varphi)\bigr)}\,.
\label{Z7}
\eeqn
The screening of the external charge--2 (adjoint) Wilson loops is
possible essentially due to the presence of the off--diagonal
field action ${\tilde S}^{\theta\varphi}$. To show this we apply the duality
transformation with respect to this action:
$$e^{ - \sum_P {\tilde S}^{\theta\varphi}_P(\theta,\varphi)}
 = \!\!\! \sum_{\stackrel{n_i  \in \Z(c_2)}{i=1,2,3}}
 \!\!\!\!\! I^H_{n_1} I^H_{n_2} I^C_{n_3}\, e^{i (H,n_1)+i (H',n_2) +i (C,n_3)}\,,$$
where the weights $I^l_{n_i}$ correspond to the terms in Eq.~\eq{S:th:phi}.

The integration of Eq.~\eq{Z7} over
$\varphi$ gives us the sum over the particle trajectories with the
charge 2:
\beqn
\cint{\varphi} \, e^{ - \sum_P {\tilde S}^{\theta\varphi}_P(\theta,\varphi)}
 = \ClosedSum{j}{1}
w(j)\,\, e^{ 2 i (j,\theta)}\,,
\label{j:sum}
\eeqn
where $w(j)$ is the weight corresponding to the ensemble of the
Wilson loops $\{j\}$. The integration over the field $\varphi$
provide us with constraints on the dynamical integer trajectories
$j$: the trajectories must be closed, $\delta j = 0$, due to
gauge invariance of both parts of Eq.~\eq{j:sum}.

The final step is to substitute the loop expansion~\eq{j:sum} into
the partition function~\eq{Z7} and to apply the arguments identical
to those in the charge $Q=2$ scalar particle considered earlier.
 Thus we get the screening of the charge-2 external charges
(gluons) and, as a result, the breaking of the adjoint string.

Let us next discuss the case of charged fermions corresponding to
quark fields. The fermion determinant can
be rewritten as a sum of self-avoiding Wilson loops composed of
charge $q=1$ particle\cite{karowski,hands}. Hence we get an action
similar to (\ref{Z6}) and the screening-confinement problem could
be solved similarly.

But generally an effective fermionic action becomes more
complicated and moreover it is desirable to adopt a form of action which
makes numerical calculations possible. We employ a Polynomial
Hybrid Monte Carlo Method~\cite{PHMC}.

Consider a polynomial $P_{\Npoly}[z]$ with an even degree $\Npoly$
which approaches $z^{-1}$ as $\Npoly$ increases. We get
\begin{eqnarray}
 \det[P_{\Npoly}[D]] & = &
  \det\left[
       T_{\Npoly}^\dagger[D] T_{\Npoly}[D]
     \right]
\label{eq:SplitdetD}
\end{eqnarray}
where
$T_{\Npoly}[z]$ is a polynomial with a degree $\Npoly/2$.
Hence the fermionic determinant for $N_f$ odd-number
of flavors is written using a pseudo-fermion representation:
\begin{eqnarray}
 (\det[D])^{N_f}
 & = &
  \left[
   \det[D P_{\Npoly}[D]]\right]^{N_f}  \label{det[D]} \\
\times&&\hspace{-.8cm}\int\!\!{\cal D}\phi^{\dag}{\cal D}\phi
  \exp\left[
       -\phi^{\dag}T_{\Npoly}^{\dag}[D]T_{\Npoly}[D]\phi
     \right]\,.\nonumber
\end{eqnarray}
Since $T_{\Npoly}[z]$ is a polynomial of degree $\Npoly/2$, then
\begin{eqnarray}
T_{\Npoly}^\dagger[D] T_{\Npoly}[D]=a_0
+a_1M+a_1^{*}M^{\dag}+\cdot\cdot\cdot,
\end{eqnarray}
where $M$ is a hopping parameter term in $D$. Since the higher-order terms are
irrelevant in the screening problem, we consider only the linear term
with respect to $M$. Here we adopt the Wilson fermion action for $M$
coupled to  a $U(1)$ compact gauge field as an example:
\begin{eqnarray}
M&=&\kappa\sum_{\mu=1}^4\{(1-\gamma_{\mu})
\left(
\begin{array}{cc}
e^{i\theta_{x,\mu}}&0\\
0&e^{-i\theta_{x,\mu}}
\end{array}
\right)\delta_{x+\mu,x'}\nonumber \\
+&&\hspace{-.3cm}(1+\gamma_{\mu})
\left(
\begin{array}{cc}
e^{-i\theta_{x-\mu,\mu}}&0\\
0&e^{i\theta_{x-\mu,\mu}}
\end{array}
\right)\delta_{x-\mu,x'}
\}
\end{eqnarray}
We separate the U(1) variant phase factor from the U(1) invariant
term ${\tilde \phi}_{\alpha}(x)$ of the pseudo-fermion field
$\phi_{\alpha}(x) = {\tilde \phi}_{\alpha}(x)
\exp(i\chi_{x,\alpha}\tau_3)$ ($\alpha=$Dirac indices). Regarding the
field ${\tilde \phi}$ as a constant  we get the following
mean-field types of interactions from the Wilson-fermion action:
\begin{eqnarray}
&&\sum_{\alpha, \mu, x}\cos(\Delta_{\mu}\chi_{x,\alpha}+\theta_{x,\mu})\,,
\nonumber\\
&&\hspace{-.3cm}\sum_{\alpha, \beta, \mu, x}\hspace{-.2cm}
\cos(\chi_{x+\mu,\beta}
-\chi_{x,\alpha}+\theta_{x,\mu})\bar{\gamma}_{\alpha\beta}(\mu)\,,
\nonumber
\end{eqnarray}
where $\bar{\gamma}(\mu)$ are realized Euclidean gamma
matrices, {\it i.e.}, $\bar{\gamma}(\mu) = i \gamma_{\mu}$ for $\mu=1,3$
and $\bar{\gamma}(\mu) = \gamma_{\mu}$ for $\mu=2,4$.

As an  effective fermionic action, it is natural to assume
the following periodic pseudo-fermion action:
\begin{eqnarray}
e^{-S_f(\theta,\chi)}
&=&e^{F(\Delta_{\mu}\chi_{x,\alpha}+\theta_{x,\mu})
+G(\chi_{x+\mu,\beta}-\chi_{x,\alpha}+\theta_{x,\mu})
\bar{\gamma}_{\alpha\beta}(\mu)}\nonumber\\
&=&\sum_{j_{\alpha,\mu}(x)\in Z}\sum_{j_{\alpha\beta,\mu}(x)\in Z}
\tilde{F}(j_{\alpha,\mu})\tilde{G}(j_{\alpha\beta,\mu})
\label{pseudo} \\
&\times&e^{ij_{\alpha,\mu}(x)(\Delta_{\mu}\chi_{x,\alpha}+\theta_{x,\mu})}
\nonumber \\
&\times&e^{ij_{\alpha\beta,\mu}(x)(\chi_{x+\mu,\beta}
-\chi_{x,\alpha}+\theta_{x,\mu})\bar{\gamma}_{\alpha\beta}(\mu)}
\nonumber
\end{eqnarray}
where $F$ and $G$ are periodic functions. In Eq.~\eq{pseudo} we have
used the Fourier (duality) transformation. The integer current
$j_{\mu}(x)\equiv\sum_{\alpha}j_{\alpha,\mu}(x)+
\sum_{\alpha,\beta}j_{\alpha\beta,\mu}(x)
\bar{\gamma}_{\alpha\beta}(\mu)$ is coupled to the field $\theta$
and carries unit electric charge.
Due to the gauge invariance of the fermion determinant
the integration of Eq.~\eq{pseudo} over the phases $\chi$ gives us a
conservation constraint for the currents, $\delta j = 0$.
Thus
\beqn
\cint{\chi} \, e^{ - S_f(\theta,\chi)}
= \ClosedSum{j}{1}
w_F(j)\,\, e^{i (j,\theta)}\,,
\label{j:sum:F}
\eeqn
where $w_F$ is a weight for the electric particle trajectories.

With the  gauge field action (\ref{Z1}), we can see that the
screening and the string breaking in full QCD also is successfully
explained  as done in the case of charged scalar particle in the
framework of $U(1)$ theory. It is again stressed that the area-law
behavior is seen in the intermediate regions even in full QCD.
Such an area-law behavior can be explained by the dual Meissner
effect due to the monopole condensation. Confinement is important
also in full QCD.

\begin{acknowledgments}
The authors are grateful to M.~I.~Polikarpov for useful
discussions. M.~N.~Ch. is supported by  JSPS
%Grant-in-Aid for JSPS Fellows
fellowship P01023 and T.~S. is supported by JSPS Grant-in-Aid for
Scientific Research on Priority Areas 13135210.
\end{acknowledgments}


\begin{thebibliography}{99}
\bibitem{'tHooft:1981ht}
G.~'t Hooft, Nucl.\ Phys.\ B {\bf 190}, 455 (1981).
%%CITATION = NUPHA,B190,455;%%
\bibitem{suzuki90}
T.~Suzuki and I.~Yotsuyanagi,  Phys.\ Rev.\ D{\bf 42}, 4257 (1990);
%%CITATION = PHRVA,D42,4257;%%
T.~Suzuki,  in {\it Continuous Advances in QCD 1996}
(World Scientific, 1997),  p. 262.
\bibitem{shiba95}
H.~Shiba and T.~Suzuki,  Phys.\ Lett.\ B{\bf 351}, 519 (1995).
%%CITATION = HEP-LAT 9408004;%%
\bibitem{nakamura}
S.~Kato  {\it et al.},
Nucl.\ Phys.\ B {\bf 520}, 323 (1998).
%%CITATION = NUPHA,B520,323;%%
\bibitem{fujimoto}
S.~Fujimoto, S.~Kato and T.~Suzuki, Phys.\ Lett.\ B {\bf 476}, 437 (2000).
%%CITATION = HEP-LAT 0002006%%
\bibitem{chernodub00}
M.~N.~Chernodub {\it et al},
Phys.\ Rev.\ D{\bf 62},  094506 (2000).
%%CITATION = HEP-LAT 0006025;%%
\bibitem{StringBreaking}
V.~G.~Bornyakov {\it et al}, talk at
Lattice 2002, Cambridge, USA, June, 2002;
C.~W.~Bernard {\it et al.}, Phys.\ Rev.\ D {\bf 64},
074509 (2001);
%%CITATION = HEP-LAT 0103012;%%
A.~Duncan, E.~Eichten and H.~Thacker,
Phys.\ Rev.\ D {\bf 63}, 111501 (2001);
%%CITATION = HEP-LAT 0011076;%%
B.~Bolder {\it et al.},
Phys.\ Rev.\ D {\bf 63}, 074504 (2001);
%%CITATION = HEP-LAT 0005018;%%
H.~D.~Trottier,
Phys.\ Rev.\ D {\bf 60}, 034506 (1999);
%%CITATION = HEP-LAT 9812021;%%
C.~DeTar {\it et al.}, Phys.\ Rev.\ D
{\bf 59}, 031501 (1999).
%%CITATION = HEP-LAT 9808028;%%
\bibitem{greensite}J.~Ambj\o rn  {\it et al}, JHEP  {\bf 0002}, 033 (2000).
%%CITATION = HEP-LAT 9907021;%%
\bibitem{ZNpicture}G.~'t Hooft, Nucl.\ Phys.\ B {\bf 138}, 1 (1978).
%%CITATION = NUPHA,B138,1;%%
\bibitem{DiffLattice}A.~H.~Guth, Phys.\ Rev.\ D {\bf 21}, 2291 (1980);
%%CITATION = PHRVA,D21,2291;%%
see also the review by
M.~N.~Chernodub and M.~I.~Polikarpov, in {\it Confinement, Duality, and
Nonperturbative Aspects of QCD}, Ed. by Pierre van Baal (Plenum, New
York 1998), p. 387.
%%CITATION = HEP-TH 9710205;%%
\bibitem{Faber} M.Zach, M.Faber, P.Skala, Nucl.Phys.B {\bf 529}, 505 (1998).
%%CITATION = HEP-LAT 9709017;%%
\bibitem{EihornSavit} M.B.Einhorn and R.Savit,
Phys. Rev. D {\bf 17}, 2583 (1978);
%%CITATION = PHRVA,D17,2583;%%
{\bf 19}, 1198 (1979);
%%CITATION = PHRVA,D19,1198;%%
M.~N.~Chernodub and M.~I.~Polikarpov, unpublished (1997).
\bibitem{degrand}T.~A.~DeGrand, D.~Toussaint,  Phys. Rev. D {\bf 22},
2478 (1980).
%%CITATION = PHRVA,D22,2478;%%
\bibitem{Chernodub:pw} M.~N.~Chernodub, M.~I.~Polikarpov and A.~I.~Veselov,
Phys.\ Lett.\ B {\bf 342}, 303 (1995).
%%CITATION = HEP-LAT 9408010;%%
\bibitem{karowski} M.~Karowski, R.~Schrader and H.~J.~Thun,
Comm.\ Math.\ Phys.\  {\bf 97}, 5 (1985).
%%CITATION = CMPHA,97,5;%%
\bibitem{hands} S.~Hands and J.~B.~Kogut,
Nucl.\ Phys.\ B {\bf 462}, 291 (1996).
%%CITATION = HEP-LAT 9509072;%%
\bibitem{PHMC} Ph.~de~Forcrand and T.~Takaishi, Nucl. Phys. B(Proc. Suppl.)
{\bf 53}, 968 (1997);
%%CITATION = HEP-LAT 9608093;%%
R.~Frezzotti and K.~Jansen, Phys. Lett. B{\bf 402}, 328 (1997);
%%CITATION = HEP-LAT 9702016;%%
C.~Alexandrou {\it et al.}, Phys. Rev. D{\bf 60}, 034504 (1999);
%%CITATION = HEP-LAT 9811028;%%
S.~Aoki  {\it et al.}, hep-lat/0112051.
%%CITATION = HEP-LAT 0112051;%%

\end{thebibliography}
\end{document}